\documentclass[3p,times,twocolumn]{elsarticle}

\usepackage{ecrc}

\volume{00}

\firstpage{1}

\journalname{Nuclear Physics B Proceedings Supplement}

\runauth{A Yu Smirnov}

\jid{nuphbp}

\jnltitlelogo{Nuclear Physics B Proceedings Supplement}

\usepackage{amssymb}

\usepackage[figuresright]{rotating}

\newcommand{\be}{\begin{equation}}
\newcommand{\ee}{\end{equation}}
\newcommand{\bea}{\begin{eqnarray}}
\newcommand{\eea}{\end{eqnarray}}

\begin{document}

\begin{frontmatter}



\dochead{}

\title{Neutrino 2012: Outlook - theory}

\author[label1]{A.~Yu.~Smirnov}
\address[label1]{International Center for Theoretical Physics, Trieste, Italy}

\begin{abstract}

Ongoing developments in theory and phenomenology are related to the measured 
large value of 1-3 mixing
and indications of significant deviation
of the 2-3 mixing from maximal one. ``Race'' for the mass hierarchy
has started and there is good chance that multi-megaton
scale atmospheric neutrino detectors with low threshold (e.g. PINGU) 
will establish the type of hierarchy. 
Two IceCube candidates of the PeV cosmic neutrinos 
if confirmed, is the beginning of new era of high energy neutrino astronomy.   
Accumulation of data on solar neutrinos 
(energy spectrum, D-N asymmetry, value of $\Delta m^2_{21}$) may uncover some
new physics. The Tri-bimaximal mixing is disfavored and  the existing discrete symmetry 
paradigm may change. The confirmed QLC prediction,   $\theta_{13} \approx
\theta_{C}/\sqrt{2}$, testifies for GUT, 
seesaw and some symmetry at very high scales. However, the same value of 
1-3 mixing can be obtained in various ways which have different implications.  
The situation in lepton sector changes from special 
(with specific neutrino symmetries, {\it etc.}) to normal,   
closer to that in the quark sector.
Sterile neutrinos are challenge for
neutrino physics but also opportunity with many interesting phenomenological
consequences. Further studies of possible
connections between neutrinos and the dark sector of the Universe
may lead to breakthrough both in particle physics
and cosmology.

\end{abstract}

\end{frontmatter}


\section{Introduction}
\label{sec:introduction}

Kyoto~\footnote{Talk given at the XXV International Conference on Neutrino 
Physics and Astrophysics, June 3 - 9, 2012, Kyoto, Japan}
 is a special 
place, full of spiritual motions, enlightening and insights 
(see e.g., the poster of the conference), 
the place to meditate and look for the signs of future.

The standard 3 neutrino framework is the reference point. 
Global view from the global fits can be 
summarized in the following way: 

1. There are three  neutrinos  
with two salient and probably related features: 
(i) smallness of the neutrino  masses, (ii) 
peculiar pattern of the lepton mixing 
which substantially  differs from the quark mixing pattern. 

2. The nature neutrino mass is  
among still missing elements.  
It  has two aspects: 
(1) Dirac versus Majorana and 
(2) ``hard'' versus ``soft''. The usual hard masses 
are generated at the electroweak 
and higher mass scales. 
The soft (enviroment dependent) contributions to the mass, 
and in general, to the  dispersion relation 
~\footnote{Recall that it is 
the dispersion relation which is probed in oscillations.} 
are due to some new interactions of neutrinos with medium 
composed of known or new fields. 

3. The main challenge for this picture is possible 
existence of sterile neutrinos. In fact, introduction of 
``steriles'' with properties required by the LSND, MiniBooNE,  
reactor and Gallium anomalies is not a small perturbation 
of the standard framework. 

Results of the global fit of the oscillation data  
updated after the conference are given  in 
\cite{global}.  
The key points, which have far going implications 
for theory, are 

\begin{figure}
\begin{center}
\vskip -0.5cm
        \resizebox{\linewidth}{!}{\includegraphics{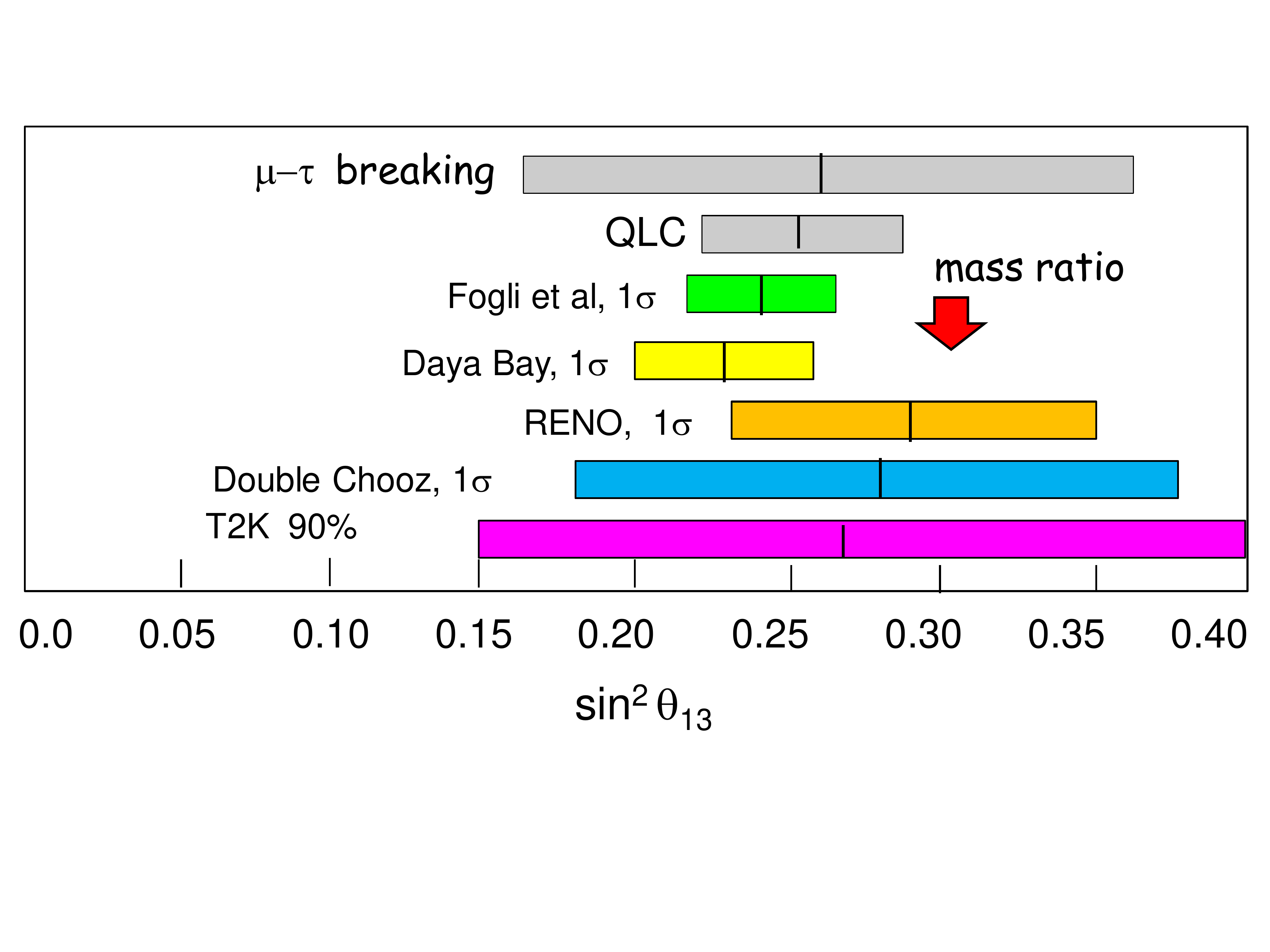}}
\vskip -1.5cm
        \caption{Determination of the 1-3 mixing. Shown are
the results from Daya-Bay experiment, T2K and global fit. 
The upper bands correspond to the QLC predictions and 
prediction from $\nu_\mu - \nu_\tau$ symmetry breaking. 
The arrow indicates the value of  mass ratio 
$\Delta m_{21}^2/\Delta m_{31}^2$.  
}
        \label{fig:sin13}
\end{center}
\end{figure}

(i) Rather large value of the 1-3 mixing angle  
(fig.~\ref{fig:sin13}). (ii) Evidence of substantial 
deviation of the 2-3 mixing from maximal one. 
(iii) Indication (from the global fit) of  preferable value 
of the CP-phase $\delta \approx 1.1 \pi$. 

Measurement of the 1-3 mixing has strong 
impact on many areas of neutrino physics \cite{minakata}.
It fixes the resonance structure of  
oscillograms of the Earth at high energies $ > 1$ GeV, 
and therefore determines  effects  for accelerator and atmospheric neutrino 
fluxes in this range. The 1-3 mixing is the key parameter  
for supernova neutrino conversion. It also affects the solar neutrinos and 
predictions for the double beta decay.  
The 1-3 mixing is the door to determination of the 
mass hierarchy and CP-violation. It has far-going theoretical 
implications. 

\section{Race for the mass hierarchy}

Determination of the neutrino mass hierarchy 
will probably be  the next step in reconstruction 
of neutrino mass and flavor spectrum. 
Physical consequences of the  hierarchy change  
can be seen immediately from the 
mass and flavor spectra. 
In the double beta decay the effective mass 
$m_{ee}^{IH} > m_{ee}^{NH}$ for the lightest mass 
$m_{lightest} < 0.05$ eV. 
In the case 
of hierarchical spectrum the sum of masses 
(probed in Cosmology) is two time larger for inverted hierarchy \cite{hcosm}: 
\be
\sum_i m_i^{IH}  =  2\sum_i m_i^{NH}  \approx  2 \sqrt{\Delta m^2_{atm}}.  
\ee 
Recall that the mass eigenstates can be marked by their 
$\nu_e$ content ({\it i.e.},  amount of admixture of the electron flavor). 
$\nu_1$ has the largest $\nu_e$ admixture, $|U_{e1}|^2$,  
$\nu_2$ - about 2 times smaller, $|U_{e2}|^2$, and 
$\nu_3$ - the smallest one $|U_{e3}|^2$. 
Consequently,  vacuum oscillations due to the 3-1 mass splitting will have 
2 times larger amplitude: $D_{31} = 4|U_{e3}|^2 |U_{e1}|^2$,   
than the amplitude due to the 3-2 mass splitting,
$D_{32} = 4|U_{e3}|^2 |U_{e2}|^2$. 
In the case of NH the frequencies of oscillations,  
$\omega_{ij} \equiv \Delta m^2_{ij}/2E$,  obey inequality
$\omega_{31} > \omega_{32}$, whereas in the case of inverted 
hierarchy:  $\omega_{31} < \omega_{32}$. 
Therefore  spectral analysis of the energy distribution 
of the electron (anti)neutrino events in the detector 
should reveal two peaks with frequencies $\omega_{31}$ and 
$\omega_{32}$. In the case of NH 
lower frequency will have smaller amplitude, whereas 
for  IH the lower frequency will have larger 
amplitude~\cite{petcov}. 
Due to finite energy resolution two peaks merge, 
and the problem will be to establish whether the 
shoulder (smaller peak) is on the left or the right hand side of the main peak.  

Matter effect makes the $\nu_e$ flavor heavier and therefore 
changes two (effective) mass spectra in matter (for NH and IH)  differently. 
The mixing in matter and resonance condition are 
determined by combination 
$|\Delta m^2_{31} \cos 2\theta_{13} - 2 EV|$. 
The change of sign of  $\Delta m^2_{31}$
is equivalent to $V \rightarrow - V$, and  
the latter is realized by transition from neutrino 
to antineutrino. Thus,  in the $2\nu-$ case 
$ NH \leftrightarrow IH$ is  equivalent 
to $\nu \leftrightarrow \bar{\nu}$.  
In particular, the resonance (resonance enhancement of 
the $\nu_e \rightarrow \nu_{\mu, \tau}$ oscillations) occurs 
in the neutrino channel for NH and 
in the antineutrino channel for IH.   

Summarizing,  there are three different approaches to 
determine the mass hierarchy: 

1. Explore matter effects on the  1-3 mixing. 
This will be possible studying the atmospheric neutrino 
fluxes with magnetized detectors, e.g. ICAL at INO \cite{ical},  or 
with huge atmospheric neutrino detectors, e.g., PINGU. 
Another possibility is the LBL  experiments:  
from NO$\nu$A \cite{nova} to  ``ultimate'' LBL proposal,  
Fermilab - PINGU \cite{winter} 
in which neutrinos will cross the core of 
the Earth and therefore will undergo the parametric 
enhancement of oscillations. 
 
Supernova neutrinos  
are sensitive to the type of mass hierarchy via 
influence of matter of a star and the Earth on the 1-3 and 1-2 mixings.   
Various effects can be used~\cite{dighe}:  

a). Strong suppression of the neutronization peak 
in the case of NH: since $\nu_e  \rightarrow \nu_3$, 
the survival probability  $P = \sin^2 \theta_{13} \approx 
0.02$, as compared to  $P = \sin^2 \theta_{12} \approx 0.31$ 
in the case of IH. 

b). Modification of the spectra of electron (anti) neutrinos 
at the accretion and cooling phases:  
the spectra become two-component, that is, 
mixture of the original  $\nu_e$ and $\nu_\mu$ spectra  
(and similarly $\bar{\nu}_e$ spectra). 
Precise composition depends on mass hierarchy. 

c).  Oscillatory modulation 
of the energy spectra due to the Earth matter effect 
should be seen in the antineutrino channel for NH,  
and in the neutrino channel  for IH. So, if the Earth matter effect 
is observed for $\bar{\nu}_e$, the NH is established. 

The problem here is that  the difference 
of fluxes of the electron and non-electron 
antineutrinos, and consequently,  
the oscillation effects in the antineutrino channel,  
which is most suitable for detection,  
are small. Furthermore,  the collective oscillations 
in inner parts of a star may 
complicate the signatures.  

2. Precise measurements of 
$\Delta m^2$ (spectral analysis of  the oscillation effects discussed above)  
with the reactor antineutrinos. This, however,  put very serious 
experimental requirements.

3. Study of observables sensitive to the 
absolute mass scale: the cosmological data and the neutrinoless double beta 
decay. 

Interestingly, existence of sterile 
neutrinos with the eV scale mass  
opens up another possibility to establish 
the hierarchy among active neutrinos: 
in this case the active-sterile neutrino oscillations  
in matter of the Earth depend on the hierarchy 
\cite{soeb-dc}.

\begin{figure}
\begin{center}
        \resizebox{\linewidth}{!}{\includegraphics{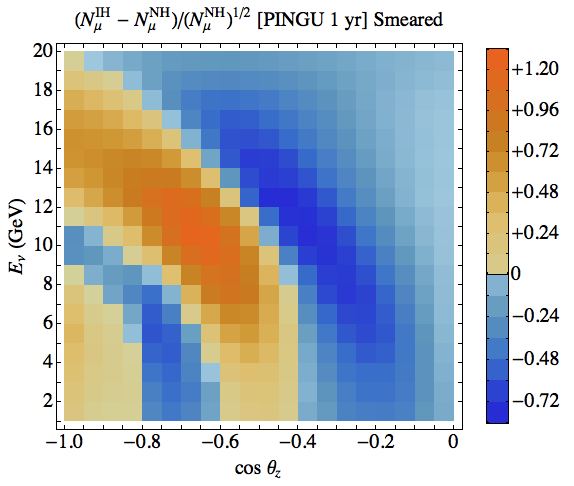}}
        \caption{Statistical significance of the determination
of the mass hierarchy. The accuracies of the
neutrino energy and angle reconstructions are taken to be
$\sigma_E =0.2 E_\nu$ and
$\sigma_\theta = \sqrt{m_p}/E_\nu$. From \cite{s-pingu}. 
}
        \label{fig:pingu}
\end{center}
\end{figure}

It may happen that next advance in the field 
will be due to huge atmospheric neutrino 
detectors (HAND's): tenth-megaton mass  scale ice (water) cherenkov 
detectors with relatively low, $\sim$ few GeV, energy threshold. 
These detectors will collect about   
$10^5$ events/year in the  range 2 - 20 GeV which 
covers the 1-3 resonance region. To determine the hierarchy it is enough 
to reconstruct the neutrino energy and direction 
with rather modest accuracy and make certain kinematical 
($E$- zenith angle) selection 
of events to avoid strong averaging.  
Fig.~\ref{fig:pingu}  illustrate sensitivity to the mass hierarchy: 
shown is the binned distribution of the quantity 
$(N_\mu^{IH} - N_\mu^{NH})/\sqrt{N_\mu^{NH}}$, 
which reflects the statistical significance of the 
determination in the $E_\nu - \cos \theta_z$ plane  
for the $\nu_\mu$ events. 
Here $N_\mu^{NH}$ ($N_\mu^{IH}$) are the number of events 
for the normal (inverted) hierarchy \cite{s-pingu}. 
This could be the fastest and 
the cheapest way to establish the hierarchy,  
although some instrumental  
(technological) developments may  be needed. 
HAND's are also sensitive to $\Delta m_{32}^2$ and 
$\theta_{23}$. Once the hierarchy 
is established one can address the issue of CP-violation searches with  HAND's.  

\section{Other missing elements}

\subsection{Nature of neutrino mass}

With EXO-200 result on the double beta decay \cite{exo200}, the  tests of 
the Heidelberg-Moscow claim \cite{h-m} 
enter critical phase.
The negative EXO-200 result,  $m_{ee} < 0.14- 0.38$ eV, 
could be in agreement with the H-M claim  
only for the smallest possible values of the 
nuclear matrix elements \cite{exo200}. 
Further increase of  statistics  and therefore  
improvements of the bound are expected soon. 
GERDA  will publish their first results in 2013 \cite{gerda}.    
Determination of the 1-3 mixing does not 
modify substantially the predicted   
regions for $m_{ee}$ in the standard $3\nu$ framework~\cite{vissani}, 
since $\theta_{13}$ is close to maximal possible value which
has been taken in computations of the regions before 
the  $\theta_{13}$ determination. 
 
Recent analysis of the cosmological data 
which includes WMAP 7 years data and 
new determinations of Hubble constant 
gives bounds on the sum of neutrino masses 
at the level $(0.3 - 0.4)$ eV ($95 \%$ C.L.) 
\cite{verde-c}. Future progress will be related to 
the next year Planck release of cosmological data. 
Although it is not clear whether  this 
will improve the bound on $\sum_i m_i$ \cite{verde-c}. 

\subsection{Deviation of the 2-3 mixing from maximal}

\begin{figure}
\begin{center}
\vskip -1cm
        \resizebox{\linewidth}{!}{\includegraphics{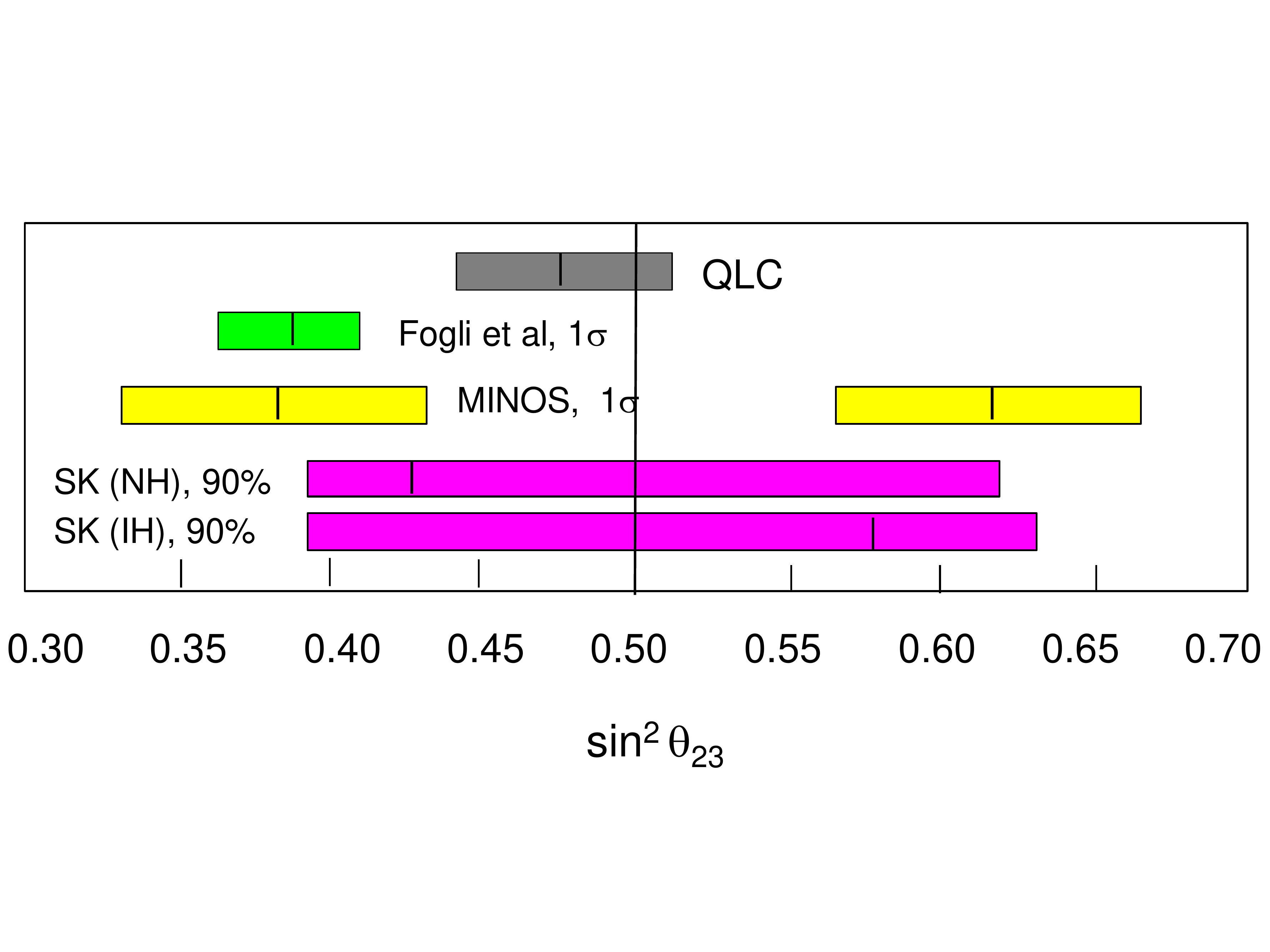}}
\vskip -1.3cm
        \caption{Determination of the 2-3 mixing. Shown are results
from MINOS experiment, Atmospheric neutrino studies and
global oscillation fit.
The arrow indicates the QLC prediction.
}
        \label{fig:sin23}
\end{center}
\end{figure}

Recent data indicate significant deviation of the 2-3 mixing 
from maximal:
\be
d_{23} \equiv 0.5 - \sin^2 \theta_{23} \sim 0.1, 
\ee
see fig.~\ref{fig:sin23}.  
In particular, direct measurements of $\sin^2 \theta_{23}$ 
by MINOS show the deviation \cite{minos23}. 
For the first time analysis of the atmospheric neutrino data
by the SK collaboration shows the deviation (with quadrant 
which depends on the mass hierarchy). 
Global  analysis  \cite{global} gives the strongest deviation. 
Recall that sensitivity to the deviation follows from 
the atmospheric neutrino data where  
the fluxes of electron neutrinos 
$F_{e}/F_e^0 - 1$ are proportional to  
$(r c_{23}^2  -  1)P_{e2}$  and 
$(r s_{23}^2  -  1)P_{e3}$
for sub- and multi-GeV events correspondingly.  
Since $r \equiv F_\mu^0 /F_e^0 \approx 2$ at low energies,   
the excess of e-like events in the sub-GeV range 
is proportional to $d_{23}$.

The appearance probability $P_{\mu e} \propto  
\sin^2 \theta_{23} \sin^2 2\theta_{13}$. So, after 
precise determination of the 1-3 mixing with reactors, 
T2K and MINOS results can be used to measure 
$\sin^2\theta_{23}$, and it is here, where sensitivity 
of LBL experiments to the 2-3 mixing comes from. 
Disappearance experiments are sensitive to 
 $P_{\mu \mu} \propto \sin^2 2\theta_{23}$ 
(T2K, MINOS). In future HAND's  (PINGU \cite{pingu}), 
may have good sensitivity to the deviation. 

The deviation is the key probe of 
the underlying physics: it (i) reflects  violation of the 
$\nu_\mu - \nu_\tau$ symmetry  as well as symmetry behind  
the TBM mixing, (ii) is probably connected to 
the non-zero 1-3 mixing; (iii) is important for the quark-lepton 
complementarity for which in the lowest order the deviation 
is small: $\theta_{23} \sim \pi/2 - V_{cb}$. 

\subsection{CP-phase: measurements and predictions}

For both mass hierarchies the global fit gives  
first glimpses of value of the CP-phase:
$\delta_{CP} \sim 1.1 \pi$ with $1\sigma$ range  $(0.4 - 1.2)~\pi$,  
but at $2\sigma$ level any value of the phase is allowed \cite{global}. 
The sensitivity to $\delta_{CP}$ comes mainly from 
the atmospheric neutrino data: the excess of the sub-GeV 
e-like events. The predicted excess can be enhanced by interference 
term for $\cos \delta_{CP} = -1$~\cite{global}.  
Future measurements will be based on comparison of 
effects in the $\nu$ and $\bar{\nu}$ channels 
(neutrino-antineutrino asymmetry) or on measurements 
of dependence of the oscillation probabilities  
in the wide energy range. The third possibility 
is reconstruction of the unitarity triangle which 
requires measurements of the depths of neutrino 
oscillations due to the solar and atmospheric 
mass splittings in the $\nu_\mu - \nu_\mu$ 
survival probability \cite{utriangle}. 

Now it is time to make predictions for $\delta_{CP}$. 
In fact, in various contexts the maximal violation value, $\delta_{CP} = \pi/2$,  
often appears (see, e.g.,  \cite{yanagida}). 
The doubt is that there is no really convincing 
theory for  the phase in the quark sector. How then the prediction 
can be made in the lepton sector where more things (e.g. see-saw 
structure) are involved?

\section{Phenomenology}

During the meeting the Venus has 
crossed the solar disc. 
That was a sign to look at the solar neutrinos again. 
And indeed, something interesting happens here.  
Still no upturn  of the energy spectrum of electrons 
has been found at low energies. The 
upturn expected according 
to LMA MSW is disfavored  at $(1.1 - 1.9)\sigma$ level \cite{smy}. 
By itself this is not very significant, however,  
no one other experiment has detected  the upturn. Furthermore, SNO \cite{sno}, 
BOREXINO \cite{borexino}, and 
KamLAND (solar) \cite{kamland-s} indicate that spectrum turns down below 
6 MeV.  Also indirectly, the  Homestake low rate supports this.
On the other hand measurements of the Be- 
and pep- neutrino fluxes by BOREXINO are 
in a good agreement with the LMA MSW prediction. 
This means (after BOREXINO) that something non-standard happens  
in the range (2 - 7) MeV. One possibility is a very light 
sterile neutrino with $\Delta m^2_{14} = (1 - 2)~ 10^{-5}$ eV$^2$,    
which mixes weakly, $\sin^2 2 \alpha = (1 - 3)~10^{-4}$,  
with $\nu_e$ \cite{pedro}.  
For hierarchical spectrum the corresponding mass  
equals $m_4 \sim (3 - 4)~ 10^{-3}$ eV.  
It may appear as a combination 
 $M^2/M_{Pl}$ with $M = (2 - 3)$ TeV which in turn, implies  
new physics at the terrascale. If the sterile  neutrino mixes 
in the state $\nu_3$ with relatively large mixing  
$\sin^2 2 \beta \approx 0.1$, an additional 
radiation in the Universe,  $\Delta N_\nu$ upto 0.9, can be 
generated.  

Another possibility is existence of the non-standard 
interactions which can modify the energy dependence 
of conversion in the intermediate energy region \cite{palazzo}. 

SK reported increasing tension  ($1.3 \sigma$ now) 
between $\Delta m_{21}^2$ measured by KamLAND 
and extracted from the analysis of solar neutrinos.
In fact,  this can be related to the absence of upturn.  
SK sees the Day-Night asymmetry of signal 
at $2.3 \sigma$ level which 
is a bit larger than LMA MSW prediction. This  
also implies larger $\Delta m_{21}^2$.
 
Solar abundance problem is still unresolved. 
There is degeneracy of effects    
of metallicity and opacity: helioseismology can not disentangle them 
\cite{serenelli}. Measurements of the CNO neutrino fluxes may help. 
These problems will be addressed  
by SNO+  \cite{mcdonald} and in future by new high statistics and 
high precision experiments HyperKamiokande, 
LENA, MICA \cite{sullivan}.

Era of  oscillation physics with huge atmospheric neutrino detectors 
has begun. 
ANTARES reported observation of the  $\nu_\mu$ oscillations 
with energy threshold $E \sim 20$ GeV  at $2.7 \sigma$ level  
\cite{antares}. DeepCore 
observes oscillation effect at energies 10 - 100 GeV 
at about $5\sigma$. 
At the same time IceCube does not see  oscillations  
at higher energies $E > 100$ GeV in agreement with standard 
oscillation predictions.  
This is important test of theory of oscillations.
It gives bounds on non-standard interactions, 
Lorentz violation, {\it etc.}.  

Supernova neutrinos: with measured value of the 1-3 mixing 
the level crossing in the H- (high density) resonance 
is highly adiabatic. This rejects many  
possibilities for flavor evolution, and picture of  
conversion becomes very simple. 
Adiabaticity is broken in shock wave fronts 
and therefore the shock waves effects should be observable. 

This picture can be affected by the collective oscillation effects 
which happen in an inner regions of a star \cite{baha}. 
The effects are more important for IH and should be realized 
during phases when neutrino density becomes 
larger than usual density. This, in principle, opens up another 
possibility to establish the mass hierarchy as well as 
to probe internal structure 
and  physical conditions of collapsing stars.

It was uncovered recently that the collective neutrino
Hamiltonian  is similar to the BCS pairing Hamiltonian describing 
superconductivity~\cite{invariants}.
This analogy allows one to write down the constants of motion
for each individual neutrino mode (characterized by momentum 
$p$)~\cite{invariants}.
The invariants are important tool to understand
various collective effects, 
in particular spectral splits, and to study stability of the collective evolution.

Recent developments were related to 
consideration of collective oscillations beyond one dimension  
where the multi-angle effects become important \cite{decoher}. 
Neutrinos arriving at a given space-time point from different directions 
(multi-angle effect) acquire different phases due to 
usual matter potential. This leads to decoherence 
and suppression of the collective phenomena. 

Summary of experimental situation with cosmic neutrinos 
is very simple: Auger see ``0'' events \cite{auger} and  IceCube sky is still 
dark \cite{ahlers}, actually almost dark:   
IceCube has reported (in rather modest way) observation of two candidates - 
cascades in the 1-10 PeV energy range \cite{icecube2}. This 
unexpected result can be beginning of new 
era in the field. Neutrinos 
can be from the diffuse cosmogenic 
flux formed during bright phase of the Universe. 

Zero Auger result is in agreement with expectations and 
therefore not dramatic.   
In contrast, null IceCube results have important implications.  
Strong bound on the neutrino flux  associated to gamma ray bursts, 
seriously restricts models of GRB  as sources of cosmic rays. 
The results have important implications in view of 
the neutrino - gamma -- CR connections.

\section{From special to normal?}

In view of  relatively large $\sin^2 \theta_{13}$, and some evidences of 
large deviation $d_{23}$, the key question 
is ``Symmetry or no symmetry?''

Data further indicate departure from the TBM mixing and 
violation of the $\nu_\mu - \nu_\tau$ symmetry. 
This can be illustrated by the best fit values of elements 
of the third column of the mixing matrix: 
$||U_{\alpha 3}|| = (0.15, 0.6, 0.8)$ instead of 
$(0.0, 0.71, 0.71)$ of TBM. Symmetry relations 
between the elements of mass matrix 
are broken substantially: E.g. in the case of NH 
the mass ratios equal $m_{\mu \mu}/ m_{\tau \tau} = 0.56$,  
and $|m_{e \mu}/ m_{e \tau}| = 0 \div  \infty$, 
instead of 1.  

The dominant line of thoughts  
and efforts during last 10 years was that 
certain residual symmetries are behind the 
approximate TBM mixing pattern \cite{residual}:

\begin{enumerate}

\item 

Mixing appears as a result of different 
ways of the original flavor symmetry, $G_f$,
breaking in the neutrino and charged lepton 
(Yukawa) sectors.  

\item

Symmetry is broken partially in each sector and 
residual symmetries are different for neutrinos,  
$G_{\nu}$,  and charged leptons, $G_{l}$.
The most popular flavor groups are  
$A_4$, $S_4$, $T^{\prime}$, $\Delta (27)$, $T_7$. 

\item

This difference of symmetry breakings
(difference of the flavor properties of the neutrino 
and charged lepton mass terms) is related 
to the Majorana nature of neutrino  
or/and  different flavor prescriptions  
for the RH components of neutrinos and  charged  
leptons. Correspondingly, different Higgs multiplets 
(flavons) participate in generation of their masses.  

\end{enumerate}
 
However, no convincing realization of this program 
has been proposed so far, although the simplest possibilities 
have been  systematically checked.  Specific models are based 
on many {\it ad hoc} assumptions, require introduction of auxiliary  
symmetries and new parameters, {\it etc.},  and usually do not lead to 
testable predictions.   
Already this posed doubts in the approach and     
new experimental results reinforced them. 
In this situation: 

1. One can further follow  the  approach 
(1 - 3) exploring various possibilities 
to accommodate recent experimental results:   
Introduce large corrections from the charged lepton sector 
which do not obey the symmetry $G_\nu$, or 
break $G_\nu$ in the neutrino sector immediately: 
$G_\nu \rightarrow 1_\nu$ \cite{altarelli}. 

2. One can still use the approach (1 - 3)  
considering ``discrete symmetries without TBM" \cite{withouttbm}.   
The ``symmetry building'' can be performed starting from 
symmetries of neutrino and charged lepton mass matrices 
in the mass basis. In the simplest version    
this leads \cite{dani} 
to the von Dyck groups, $G_f = D(2, m, p)$   ($m$, $p$ are integers)  
which include $A_4$, $S_4$, $A_5$, 
and gives two relations between the  elements of mixing matrix.  
This  fixes 2 out of 4 mixing parameters. 
The relations are for one of the column ($j$) of 
the mixing matrix:
\be
|U_{\beta j}|^2 = |U_{\gamma j}|^2, ~~~ 
|U_{\alpha j}|^2 = \frac{1 -a}{4 \sin^2 (\pi k/m)},  
\label{eq:relations}
\ee
where $\alpha \neq  \beta \neq \gamma $ 
are the flavor indices $j = 1, 2, 3$; 
$k \leq m$ is integer and  $a$ is determined from the conditions 
\be
\lambda^3_i + a \lambda^2_i - a^*\lambda_i -1 = 0, 
~~~~ \lambda_i^p = 1.
\ee
Here $i = 1, 2, 3$. The most interesting possibility is  $p = 4$,  $m = 3$  
which corresponds to $S_4$ group. In this case $a = -1$ and for $j = 1$ 
one has,  e.g., $|U_{e 1}|^2 = 2/3$.
For the best fit values of 
$\theta_{13}$ and $\theta_{23}$ one predicts then  
$\delta_{CP} = 103^{\circ}$.  Without 
simplifications the approach can lead to  4 relations 
thus fixing the mixing parameters completely. 

In view of the fact that also 1-2 mixing 
deviates from the TBM value one can consider 
the $\nu_\mu - \nu_\tau$ symmetry only, which  
would imply the equalities $\sin^2 \theta_{13} = d_{23} = 0$. 
Violation of this symmetry leads generically to related 
non-zero values of these parameters. 
If violation has ``universal" character
such that 
$m_{\mu \mu} - m_{\tau \tau} \approx m_{e \mu} - m_{\mu \tau}$
one obtains 
\be
\sin^2 \theta_{13} = \frac{1}{2} \cos^2 2 \theta_{23} 
= 2d_{23}^2 \approx 0.022 
\label{eq:mutau}
\ee
in perfect agreement with measurement. 

3. One can abandon both the 
symmetry approach (1 - 2) and TBM. 
The Quark-lepton complementarity (QLC) 
is another realization of special zero order 
structure.  QLC predicted \cite{qlc} 
\be
\sin \theta_{13} \approx \frac{1}{\sqrt{2}} \sin \theta_C
(1 - V_{cb} \cos \delta) - V_{ub} \approx \frac{\theta_C}{\sqrt{2}} 
\label{eq:13qlc}
\ee
where $\theta_C$ is the Cabibbo angle. 
This prediction is essentially 
result of permutation of the matrices of 
maximal 2-3 rotation and  1-2 rotation on 
the Cabibbo angle: 
$$
U_{12}(\theta_C) U_{23}\left(\frac{\pi}{4}\right) \approx
U_{23}\left(\frac{\pi}{4}\right)  
U_{13}\left(\frac{\theta_C}{\sqrt{2}}\right)
U_{12}\left(\frac{\theta_C}{\sqrt{2}}\right).     
$$ 
(Such a permutation is needed to reduce the PMNS matrix to the standard 
The simplest origin of the above structure 
is the following: In certain basis 
the matrix of up-quarks is diagonal, 
so that $V_u = I$, whereas diagonalization of 
the neutrino mass matrix gives the bi-maximal mixing 
$U_\nu = U_{12}(\pi/4)U_{23}(\pi/4)$. 
The latter may follow from the see-saw 
mechanism of neutrino mass generation. 
For the charged leptons and down-type quarks: 
$U_l \approx V_d = V_{CKM}$ due to Grand Unification or 
the same horizontal symmetry. As a result: 
\bea
U_{PMNS} & = & U_l^{\dagger} U_\nu = 
V_{CKM}^{\dagger}U_{bm}, 
\nonumber\\ 
V_{quarks} & = & V_u^{\dagger} V_d = V_{CKM}. 
\eea
Taking in the PMNS mixing matrix $V_{CKM}^{\dagger} \approx U_{12}(\theta_C)^{\dagger}$ 
we obtain the required mixing structure and 
$\sin^2 \theta_{13} = 0.5 \sin^2 \theta_C$ \cite{qlc,antush}. 

In the QLC framework the  exact bimaximal mixing leads  
to deviation of the 2-3 mixing from maximal: 
$d_{23} \approx  \cos \theta_C V_{cb} \cos \delta + 0.5 \sin^2\theta_C$  
which is about 0.06 for $\delta = 0$. 
Some corrections to the above picture may be needed. 
If neutrino mixing deviates from 
the bi-maximal one:   
$\sin \theta_{13} \approx \sin \theta_{23} \sin \theta_{C}$.   

Also  the QLC implies  special 
structure for neutrinos which gives the bi-maximal mixing. 
The latter can be a consequence of certain symmetry of the 
RH neutrino mass matrix. 

The weak complementarity \cite{weakcom} or Cabibbo ``haze'' \cite{hase}
essentially mean that there are corrections to 
zero order structure of mixing matrix 
of the order of $\theta_C$. E.g. the deviation of 2-3 mixing 
from maximal is  of this order. 
The ratio of masses can also be determined by $\theta_C$.  
The corrections are feature of the flavor 
physics and they do not imply quark-lepton symmetry and unification.  
No exact prediction for the angles can be done in this context.  

The self-complementarity \cite{selfcompl} is purely leptonic relation 
\be
\theta_{12} +  \theta_{13}  = \theta_{23}.  
\label{eq:self}
\ee
It is also reproduced by QLC.

4.  One can reconsider the quark-lepton universality which means  that there 
is nothing special in the lepton sector (apart from seesaw) 
and the Dirac leptonic mass matrices are organized in the same way as 
the quark mass matrices.  
Prediction for the 1-3 mixing was obtained from 
``naturalness'' of mass matrix \cite{akhm}:  
\be
\sin^2 \theta_{13} = A 
\frac{\Delta m_{21}^2}{\Delta m_{32}^2},    
\label{eq:natur}
\ee
where  $A = 0.78$ ($\sim 1 - \sin\theta_C$?) for the best fit value of $\theta_{13}$ 
This relation  
follows from the fact that there are two large mixing connecting neighboring 
generations, and from the following  two assumptions: (i) 
NH, (ii) absence of fine tuning between different elements 
of the mass matrix (e.g.,  $m_{e\mu} = m_{e \tau}$).   
There are also models where the relation (\ref{eq:natur}) is a consequence 
of certain symmetry \cite{eq:fri-ma}.   

A kind of Fritzsch ansatz can be used for the lepton 
Dirac mass matrices. If all RH neutrino masses are equal each other,  
this leads via seesaw to the normal mass hierarchy and correct value of 
1-3 mixing \cite{fukugita}.  

Another indication of the universality 
is that equalities of the same type  
$$
\theta_{13} \approx \frac{1}{2}\theta_{12}\theta_{23}~~  
{\rm and}~~ V_{ub} \approx \frac{1}{2} V_{us} V_{cb}
$$ 
are satisfied in the lepton and quark sectors. 

5. Completely opposite  approach is the mixing anarchy, 
in which mixing angles appear as random numbers \cite{anarchy}.  
At $1\sigma$ level it gives  $\sin^2 \theta_{13} > 0.025$. 
Fit with anarchy can also be considered as a test of complexity 
behind neutrino mass and mixing. 

Thus,  the observed value of 1-3 mixing 
is reproduced  in various ways (\ref{eq:mutau}), 
(\ref{eq:13qlc}), (\ref{eq:self}), (\ref{eq:natur}), 
which have completely different 
implications for fundamental theory. 
  
Predictions for unknown yet quantities can be obtained 
reducing number of parameters and structures  
of a  model applying the Occam's razor \cite{yanagida}. 
Good to know,  though,  where to cut: at the level of 
parameters, or assumptions,  or  
principles. Usually reduction of number of parameters 
is achieved by prize of increasing number of assumptions. 
For instance,  it was assumed in \cite{yanagida} 
that only two RH neutrinos are involved in see-saw, 
and there are two zeros in the Dirac mass matrix. This leads 
to phenomenologically viable scenario for the inverted 
mass hierarchy with its generic prediction for the 
Majorana  mass  $m_{ee} \approx 0.05$ eV 
and the CP-phase $\delta_{CP} \approx \pi/2$ \cite{yanagida}. 

Another realization of the Occam's razor, applied to  
number of assumptions is the following. 
(i) The high mass scale seesaw mechanism, 
probably associated to GUT scale, with 3 RH neutrinos explains smallness 
of neutrino mass. 
(ii) The Dirac Yukawa couplings in quark and lepton sectors 
are similar.  (iii) The same seesaw  
is responsible for enhancement of lepton mixing. 
So, difference of the quark and lepton mixings 
is due to seesaw.
(iv) Flavor structure originates from physics at very high scale, probably above GUT. 
The RH neutrino mass matrix may have a special structure. 
Some hidden sector may exist at GUT or/and Planck scales.

After many recent speculations, are we back to good old picture? 
Plausible scenario is SO(10) GUT with   
hidden sector (order of hundred SO(10) singlets). 
Fermionic components of this sector  
can mix with neutrinos only \cite{multi-rh}. It is this mixing that   
produces  all the observable differences 
in quark and lepton sectors: 
(i) smallness of neutrino mass 
via the double seesaw;   
(ii) required scale of the RH neutrinos;   
(iii) enhancement of lepton mixing; 
(iv) certain relations  between mass matrix elements 
via symmetries in the hidden sector; 
(v) the bi-maximal mixing required by QLC; 
(vi) it can also produce randomness (anarchy) in the lepton mixing. 


Another possibility is additional vector-like GUT multiplets whose 
components mix with the components of the chiral multiplets.  

Whole the  beauty of the usual seesaw is that it 
explains smallness of neutrino mass by existence 
of a high mass scale without recorring to  
suppressed Yukawa couplings. The TeV-scale versions of 
seesaw which can be tested at LHC require both 
new mass scale and smallness (or fine tuning) of the neutrino Yukawa couplings. 
These are  kind of ``searches under a lamp" without serious motivations 
beside a possibility to test it at LHC. 
One exception is scenario which explains 
the Heidelberg-Moscow result on $\beta \beta 0\nu$ decay by 
exchange of new particles of the L-R symmetric models 
(RH neutrinos, double charged Higgs bosons) \cite{goran}. 

Anyway,  according to \cite{tao} LHC can detect seesaw 
``messengers''  with masses below TeV:  
RH neutrinos of the seesaw type I  with 
$M_N \sim 10 - 400$ GeV, and 
$M_N$ up to  $400$ GeV in the L-R models;
Higgs triplet of the seesaw type-II with 
$M_\Delta = 400 - 1000$ GeV; the 
RH neutrino of the seesaw type-III with 
mass up to 800 GeV. 

Phenomenologically viable alternative elaborated  by the Occam's razor
is the $\nu$MSM \cite{numsm} with the keV - GeV scale seesaw.

Different approach to understand neutrino mass and mixing 
can be developed assuming that Yukawa couplings have dynamical origin being 
identified with  VEV's of scalar fields - flavons \cite{gavela}. 
The flavons transform under certain representations of the 
flavor group. Values of VEV's, and consequently,  Yukawa couplings 
are fixed by minimization of the potential of flavons fields which 
is invariant under the flavor group. In this context 
(as it was illustrated by a toy model) one can understand 
difference of mixings in quark and lepton sectors  
employing the Majorana nature of neutrinos.  

The hope was (as before) that studies of neutrinos  
will uncover something exiting  which shed light on other problems 
of  particle physics, astrophysics and Cosmology. Special values of lepton mixing  
with certain flavor symmetries behind, 
indeed,  supported the expectation. 
However,  recent developments put some doubts  
and it seems we return back  from ``special  to normal''.      

\section{Sterile neutrinos: challenge and opportunity}

``Maiko-san  session''  gave another sign.  
Recall that in the beginning three  Maiko-san's  dressed in colorful 
kimonos were dancing. Then  we saw 
two Maiko-sans in black and white.  
Then all five were dancing. A sign  of 3 + 2 ! 
However, in the interpretation part only three Maiko-sans showed 
up and senior lady was giving explanations.
I am not sure that this is 3 + 1. 

In any case possible existence of sterile neutrinos 
is the challenge for everything: 
theory, phenomenology and experiment. 

Theory: 
corrections to mass eigenvalues and mixing angles of active neutrinos 
from the  mixing with eV scale states required by 
LSND are, in general,  of the order one. They change structure and 
symmetries of the mass matrix of active components significantly, 
unless special conditions  are imposed on the  
active-sterile mixing (see \cite{rodejohann}). On the other hand this 
mixing can be used to explain certain properties of 
active neutrinos, e.g.,  it can enhance mixing between active 
neutrino components. So, without clarification of 
existence of the eV steriles our further progress 
in understanding neutrino mass and mixing is almost impossible.

Phenomenology: the eV steriles  are important for  
solar, atmospheric and supernova neutrinos, for  
beta and double beta decays, for reactor and accelerator 
neutrinos, for cosmology (nucleosynthesis, 
extra radiation of the universe, structure formation) \cite{whitepaper}. 
They can affect searches of the dark matter. 

\begin{figure}
\begin{center}
        \resizebox{\linewidth}{!}{\includegraphics{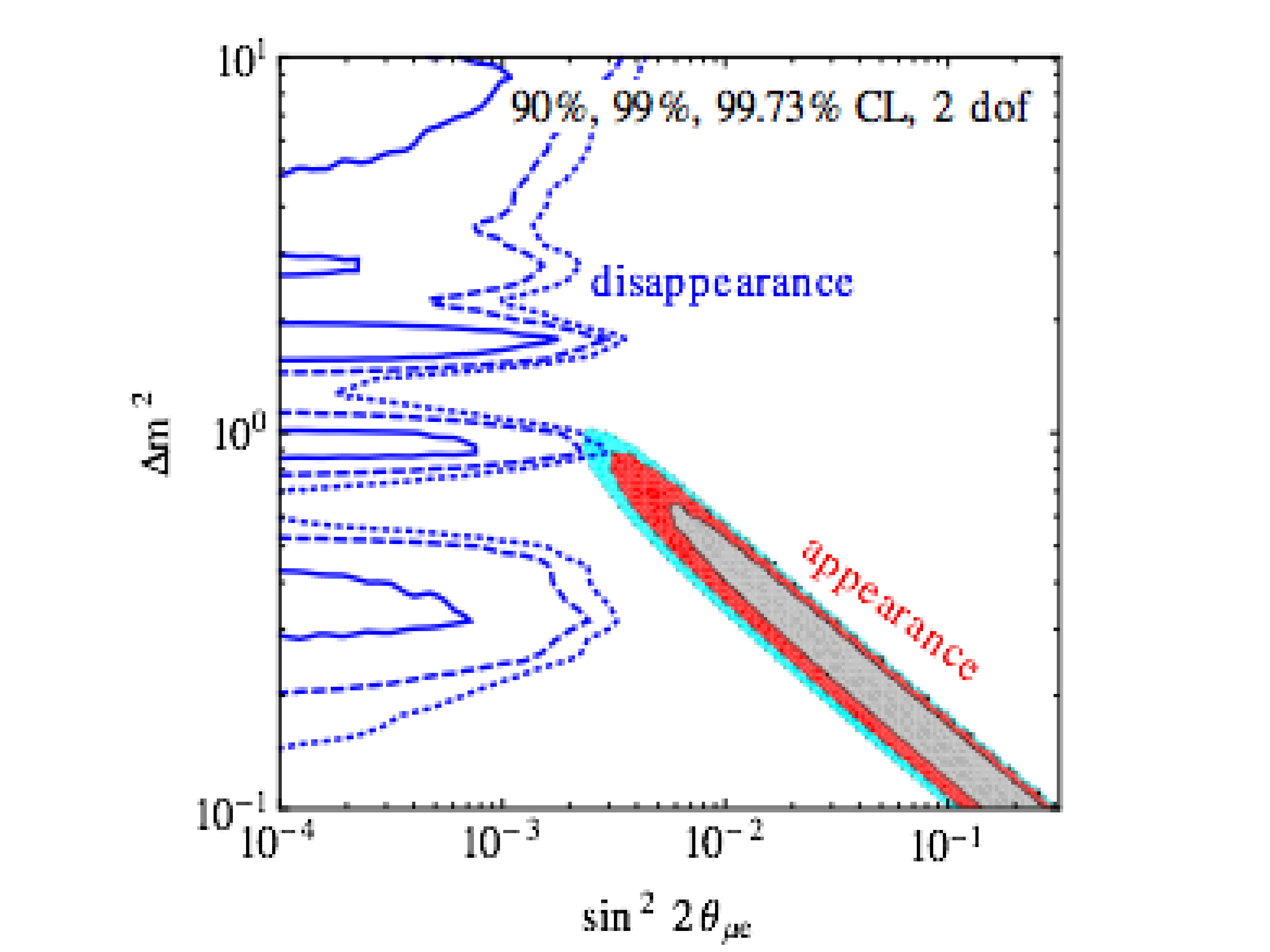}}
        \caption{
Allowed and excluded regions of parameters of sterile
neutrino $\Delta m_{41}^2 - \sin^2 2\theta_{e\mu}$ 
within 3 + 1 scheme. From \cite{globals}. }
        \label{fig:sterile}
\end{center}
\end{figure}

Experiment: situation is rather controversial.  
MiniBooNE confirms LSND: 
similar excess is observed both in 
neutrino and antineutrino channels.  
But good description of the energy dependence 
of the  MiniBooNE excesses  would be  possible with two steriles 
and CP-violation \cite{miniboone}. 
At the same time new contributions to the appearance signals 
have been revealed which reduces significance of the 
LSND (and also MiniBooNE) excess 
\cite{lsndbac}. The issues of the cross-section and energy reconstruction 
are under discussion.     
With negative MINOS searches for steriles \cite{minoss}  
tension between the disappearance data and the appearance 
LSND-MiniBooNE signals further strengthened~\cite{globals}. According to  
fig.~\ref{fig:sterile}    
there is a small region  around 1 eV$^2$ 
in which $3\sigma$ allowed and preferred regions overlap.  
With only one sterile fit of the energy spectra is far from 
being perfect.

Situation with the reactor and Gallium anomalies \cite{gallium}
is less dramatic. There is new bound obtained 
from joint fit of solar, KamLAND, Daya-Bay and 
Reno experiments: $\sin^2 2 \theta_{es} < 0.2$ \cite{globals}. 
Global fit gives the best value  
$\Delta m^2 = 1.78$ eV$^2$ which, however,  is not consistent with 
$\nu_\mu-$ oscillation results. 
Analysis of cosmological data in terms of 
$\Lambda$CDM leads to the bound 
$\Delta m^2_{41} < 0.25 $ eV$^2$ for 1 sterile \cite{cosm-s}.

For effective number of neutrino species (extra radiation 
in the Universe) one has  $N_\nu = 3.5 \pm 0.3$, $1 \sigma$ \cite{extrar}. 

IceCube is ideal detector of the eV mass steriles \cite{icecube-s}. 
For  $\Delta m^2 \sim 1 $ eV$^2$ 
the  MSW resonance is realized in the  $\bar{\nu}_\mu - \bar{\nu}_s$ 
channel in matter of the Earth at neutrino energies $\sim 1$ TeV.   
The resonant enhancement of  
oscillations in this energy range leads to distortion of the 
zenith angle distribution of the $\mu-$ like atmospheric 
neutrino events. The effect 
(which depends also on admixture 
$U_{\tau 4}$) is of the order $10 - 20 \%$. Statistics   
(few $\times 10^3$ events in each of 20 bins)  
is not a problem. Once systematic uncertainties are   
understood, IceCube will  provide critical test of 
existence of the eV steriles.

\section{Neutrinos and the Dark universe}

There are various ideas about how neutrino can be related to 
the dark sector of the Universe: dark matter and dark energy. 
Direct connection: new neutrino states with the kev scale mass  
can compose the worm DM. 
Indirect connections: 
the same flavor symmetries which are  
responsible for  mixing pattern 
or/and for smallness of neutrino mass can ensure stability of DM.  
New particles involved in mechanisms of 
neutrino mass generation can play the role of DM. 

Neutrino mass generation can be related via 
the seesaw with inflation, leptogenesis 
and production of the Dark matter particles: 
``everything in one'' [scenario] 
\cite{buchmuller}.  Indeed, the inflaton field 
can act (in SUSY context) as a driving field which leads to 
VEV  of scalar  $S$ that breaks the B-L symmetry:  
$S$ couples with the RH neutrinos, $N_i$,  and generate their masses. 
Then evolution can proceed in the following way: 
Reheating  occurs via decays of $S \rightarrow N N$ and subsequent decay 
of RH neutrinos in to SM particles: $N \rightarrow l  H$. 
The decay of the lightest of them,  $N_1$,  with mass $M_1$   
produced the lepton asymmetry which then converted to the baryon asymmetry. 
The reheating  temperature $T_{RH} = T_{RH}(M_1)$ determines the relic density 
of the thermally produced gravitinos which play the role of the DM particles.   
The successful scenario is realized for $v_{B-L} \sim 5~10^{15}$ GeV, 
$M_1 \sim 10^{11}$ GeV  and $T_{RH} \sim 10^{10}$ GeV and the effective 
light neutrino mass $\tilde{m}_1 \sim 0.04$ eV.  
The latter implies the bound on mass of gravitino 
$m_g > 10$ GeV.   

Hints of extra radiation are another driving force of 
developments.  
Still BBN prefers $N_\nu > 3$, although 
recent determination of the deuterium abundance ~\cite{lverde}
results in $N_\nu = 3.0 \pm 0.5$ \cite{verde-c}. 
The CMB data give stronger indications. 
With CMB, higher value of the Hubble constant drives 
$N_\nu$ to $3.5 - 3.7$. The highest value of 
$\Delta N_\nu$ follows from analysis which includes data from 
WMAP, ACBAR and BAO. 
High values of H0 tend to increase $\Delta N_\nu$,  
whereas ACT data decreases it \cite{melchiori}.  
The Planck cosmological data are expected to be sensitive 
to $\Delta N_\nu \approx 0.2$. 

Another possible connection: 
the neutrino velocity can be affected by the 
dark sector of the Universe. 
One still can explore whether, e.g., ``short cut in extra dimensions'' 
can be the reason of bad fiber connection in the OPERA experiments. 
Instead, I will make few  statements which could be of some relevance. 

1. Neutrino is the lightest massive particle we know. 
Therefore for available energies its velocity can be the 
closest one to velocity of light. 

2. If measurements of neutrino velocity are continued, 
an important question is what can be checked 
at the achievable level of sensitivity? 
Numerous papers issued recently (worth to analyze part of them) 
contain some answers to this. 

3. Fundamental symmetries can be violated effectively due 
to interactions  with some background. 
Recall, e.g., that  CPT is violated in oscillations 
in usual medium. 

4. In this connection, do we know well  the background?
Do we know everything about dark sector of the Universe?

5. The proposal of neutrino oscillations by B. Pontecorvo  
was motivated by rumor that Ray Davis saw effect 
in the Cl-Ar detector from  atomic reactor \cite{bilenky}.

\section{Outlook}

1. To a large extend future developments in theory and phenomenology 
will be driven by new experimental highlights:  
rather large value of $\sin^2 \theta_{13}$   
and indication of significant deviation of  
the 2-3 mixing   
from maximal. Global fit gives first glimpses on to the CP-phase,  
and it is the time to make predictions for  $\delta_{CP}$.

2. Interesting developments can be related to the solar neutrinos, where  
absence of the spectral upturn is further confirmed. That can be related to 
some tension with KamLAND value of $\Delta m^2_{21}$ and a bit higher 
value of the day-night asymmetry reported by SK. 
Is this just accidental, statistical fluctuation or 
accumulation of results which testify for new physics,  
e.g.,  new very light steriles or  new interactions?   

3. The race for  mass hierarchy  has started. Although 
phenomenology of  different hierarchies is well elaborated 
still some more ideas to identify ordering may appear. 
It may happen that new 
developments will be associated to huge atmospheric neutrino detectors (HAND's) 
with low  ($\sim 1$ GeV) energy threshold. 
HAND's may perform good measurements of $\theta_{23}$. 
Sensitive search of steriles with IceCube will be realized. 
Once hierarchy is established one can explore in more details 
searches of CP-violation. A possibility to determine CP-violation  
with HANDs is challenging but not completely excluded.   

4. In perspective, the threshold of the huge detectors  can be 
further reduced down to $10^{-2}$ GeV (MICA). Physics potential of this 
type of detection still should be explored.  

5. Presentation of the two IceCube candidates  
of cosmic neutrinos (cascades) was rather modest.  
The discovery (if confirmed) will have enormous impact on the field and  
already these two events 
trigger various speculations. Implications of the null IceCube result, in particular, 
for searches of neutrinos associated to GRB will be in the center of studies.  

6. The tribimaximal mixing is further disfavored. 
Still TBM can be considered as the lowest order structure which requires 
significant corrections. In view of the fact that no convincing model for that  
has been proposed, the paradigm may change. 
E.g. one can abandon TBM and use  some other zero order structure,   
or  pursue  the same flavor symmetry approach without TBM.  

The relation $\theta_{13} \sim \theta_C/\sqrt{2}$  
predicted in the context of QLC, is in a good agreement 
with recent measurements.  Is this accidental? 

It could be certain shift in our understanding of mass and mixing 
from  ``special to normal'', and  things with leptons become closer to quarks. 
Probably the Dirac
Yukawa structures are similar in both cases, symmetry (if exists) is  the same 
for quarks and leptons, there is no   
special symmetry in the lepton sector.   
Smallness of neutrino mass and large 
lepton mixing originate from the same seesaw.   
In this connection we expect normal mass hierarchy, 
mass matrices with flavor ordering, 
high scale seesaw,  enhancement of mixing. GUT embedding may require 
new elements, in particular, mixing of neutrinos with new singlets 
of the gauge symmetry group.

7. Steriles are challenge for neutrino physics.  
There are controversial experimental evidences, tension 
with cosmology, tension between the  appearance and disappearance data,  
puzzling theoretical situation. 
On the other hand, steriles have rich phenomenology  and 
their existence opens up  new ways to understand 
the observed mixing and masses.   
New searches of steriles with  atmospheric neutrinos will be performed using 
IceCube, DeepCore, and dedicated source, reactor and accelerator experiments.

8. Neutrinos and dark Universe will continue to be one of exiting  areas of research. 
Further studies of possible connections may 
lead to breakthrough both in particle physics and 
Cosmology.






\end{document}